\input harvmac
\let\includefigures=\iftrue
\let\useblackboard==\iftrue
\newfam\black

\includefigures
\message{If you do not have epsf.tex (to include figures),}
\message{change the option at the top of the tex file.}
\def\figin{\epsfcheck\figin}\def\figins{\epsfcheck\figins}
\def\epsfcheck{\ifx\epsfbox\UnDeFiNeD
\message{(NO epsf.tex, FIGURES WILL BE IGNORED)}
\gdef\figin##1{\vskip2in}\gdef\figins##1{\hskip.5in}
\else\message{(FIGURES WILL BE INCLUDED)}%
\gdef\figin##1{##1}\gdef\figins##1{##1}\fi}
\def\DefWarn#1{}
\def\figinsert{\goodbreak\midinsert}
\def\ifig#1#2#3{\DefWarn#1\xdef#1{fig.~\the\figno}
\writedef{#1\leftbracket fig.\noexpand~\the\figno}%
\figinsert\figin{\centerline{#3}}\medskip\centerline{\vbox{
\baselineskip12pt\advance\hsize by -1truein
\noindent\footnotefont{\bf Fig.~\the\figno:} #2}}
\endinsert\global\advance\figno by1}
\else
\def\ifig#1#2#3{\xdef#1{fig.~\the\figno}
\writedef{#1\leftbracket fig.\noexpand~\the\figno}%
\global\advance\figno by1} \fi
\def\id{{1 \kern-.28em {\rm l}}}

\def\K3{{\bf K3}}
\def\journal#1&#2(#3){\unskip, \sl #1\ \bf #2 \rm(19#3) }
\def\andjournal#1&#2(#3){\sl #1~\bf #2 \rm (19#3) }

\def\bar{\overline}
\def\hat{\widehat}
\def\ie{{\it i.e.}}
\def\eg{{\it e.g.}}

\def\tilde{\widetilde}

\def\frac#1#2{{#1\over#2}}

\def\half{\frac12}

\def\inbar{\,\vrule height1.5ex width.4pt depth0pt}
\def\IC{\relax\hbox{$\inbar\kern-.3em{\rm C}$}}
\def\IR{\relax{\rm I\kern-.18em R}}
\def\IZ{\relax{\rm I\kern-.18em Z}}

%
%

%
\catcode`\@=11
\def\slash#1{\mathord{\mathpalette\c@ncel{#1}}}
\overfullrule=0pt

\def\CC{{\cal C}}

\def\EE{{\cal E}}
\def\FF{{\cal F}}

\def\HH{{\cal H}}

\def\JJ{{\cal J}}

\def\LL{{\cal L}}
\def\MM{{\cal M}}
\def\NN{{\cal N}}

\def\ZZ{{\cal Z}}

\def\underrel#1\over#2{\mathrel{\mathop{\kern\z@#1}\limits_{#2}}}

\catcode`\@=12


%

\def\exp{{\rm exp}}


\def\ie{{\it i.e.}}
\def\eg{{\it e.g.}}


\lref\GiveonFU{
  A.~Giveon, M.~Porrati and E.~Rabinovici,
  ``Target space duality in string theory,''
Phys.\ Rept.\  {\bf 244}, 77 (1994).
[hep-th/9401139].
}

\lref\SmirnovLQW{
  F.~A.~Smirnov and A.~B.~Zamolodchikov,
  ``On space of integrable quantum field theories,''
Nucl.\ Phys.\ B {\bf 915}, 363 (2017).
[arXiv:1608.05499 [hep-th]].
}

\lref\CavagliaODA{
  A.~Cavaglià, S.~Negro, I.~M.~Szécsényi and R.~Tateo,
  ``$T \bar{T}$-deformed 2D Quantum Field Theories,''
JHEP {\bf 1610}, 112 (2016).
[arXiv:1608.05534 [hep-th]].
}

\lref\LeFlochRUT{
  B.~Le Floch and M.~Mezei,
  ``Solving a family of $T\bar{T}$-like theories,''
[arXiv:1903.07606 [hep-th]].
}

\lref\GiveonNIE{
  A.~Giveon, N.~Itzhaki and D.~Kutasov,
  ``$ T\bar{T} $ and LST,''
JHEP {\bf 1707}, 122 (2017).
[arXiv:1701.05576 [hep-th]].
}

\lref\KutasovXB{
  D.~Kutasov,
  ``Geometry on the Space of Conformal Field Theories and Contact Terms,''
Phys.\ Lett.\ B {\bf 220}, 153 (1989).
}

\lref\GiveonMYJ{
  A.~Giveon, N.~Itzhaki and D.~Kutasov,
  ``A solvable irrelevant deformation of AdS$_{3}$/CFT$_{2}$,''
JHEP {\bf 1712}, 155 (2017).
[arXiv:1707.05800 [hep-th]].
}

\lref\ChakrabortyVJA{
  S.~Chakraborty, A.~Giveon and D.~Kutasov,
  ``$ J\overline{T} $ deformed CFT$_{2}$ and string theory,''
JHEP {\bf 1810}, 057 (2018).
[arXiv:1806.09667 [hep-th]].
}

\lref\ApoloQPQ{
  L.~Apolo and W.~Song,
  ``Strings on warped AdS$_{3}$ via $ T\bar J $ deformations,''
JHEP {\bf 1810}, 165 (2018).
[arXiv:1806.10127 [hep-th]].
}

\lref\PolchinskiRQ{
  J.~Polchinski,
  ``String theory. Vol. 1: An introduction to the bosonic string,''
}

\lref\GiveonMYJ{
  A.~Giveon, N.~Itzhaki and D.~Kutasov,
  ``A solvable irrelevant deformation of AdS$_{3}$/CFT$_{2}$,''
JHEP {\bf 1712}, 155 (2017).
[arXiv:1707.05800 [hep-th]].
}

\lref\CardySDV{
  J.~Cardy,
  ``The $ T\overline{T} $ deformation of quantum field theory as random geometry,''
JHEP {\bf 1810}, 186 (2018).
[arXiv:1801.06895 [hep-th]].
}

\lref\ChakrabortyKPR{
  S.~Chakraborty, A.~Giveon, N.~Itzhaki and D.~Kutasov,
  ``Entanglement Beyond $\rm AdS$,''
[arXiv:1805.06286 [hep-th]].
}

\lref\GuicaLIA{
  M.~Guica,
  ``An integrable Lorentz-breaking deformation of two-dimensional CFTs,''
SciPost Phys.\  {\bf 5}, no. 5, 048 (2018).
[arXiv:1710.08415 [hep-th]].
}

\lref\GiveonNS{
  A.~Giveon, D.~Kutasov and N.~Seiberg,
  ``Comments on string theory on AdS(3),''
Adv.\ Theor.\ Math.\ Phys.\  {\bf 2}, 733 (1998).
[hep-th/9806194].
}

\lref\KutasovXU{
  D.~Kutasov and N.~Seiberg,
  ``More comments on string theory on AdS(3),''
JHEP {\bf 9904}, 008 (1999).
[hep-th/9903219].
}

\lref\GiveonZM{
  A.~Giveon, D.~Kutasov and O.~Pelc,
  ``Holography for noncritical superstrings,''
JHEP {\bf 9910}, 035 (1999).
[hep-th/9907178].
}

\lref\DattaTHY{
  S.~Datta and Y.~Jiang,
  ``$T\bar{T}$ deformed partition functions,''
JHEP {\bf 1808}, 106 (2018).
[arXiv:1806.07426 [hep-th]].
}

\lref\AharonyBAD{
  O.~Aharony, S.~Datta, A.~Giveon, Y.~Jiang and D.~Kutasov,
  ``Modular invariance and uniqueness of $T\bar{T}$ deformed CFT,''
JHEP {\bf 1901}, 086 (2019).
[arXiv:1808.02492 [hep-th]].
}

\lref\AharonyICS{
  O.~Aharony, S.~Datta, A.~Giveon, Y.~Jiang and D.~Kutasov,
  ``Modular covariance and uniqueness of $J\bar{T}$ deformed CFTs,''
JHEP {\bf 1901}, 085 (2019).
[arXiv:1808.08978 [hep-th]].
}

\lref\DubovskyBMO{
  S.~Dubovsky, V.~Gorbenko and G.~Hernández-Chifflet,
  ``$ T\overline{T} $ partition function from topological gravity,''
JHEP {\bf 1809}, 158 (2018).
[arXiv:1805.07386 [hep-th]].
}

\lref\ChakrabortyMDF{
  S.~Chakraborty, A.~Giveon and D.~Kutasov,
  ``$T\bar{T}$, $J\bar{T}$, $T\bar{J}$ and String Theory,''
[arXiv:1905.00051 [hep-th]].
}

\lref\HashimotoHQO{
  A.~Hashimoto and D.~Kutasov,
  ``Strings, Symmetric Products, $T \bar{T}$ deformations and Hecke Operators,''
[arXiv:1909.11118 [hep-th]].
}

\Title{} {\centerline{$T \bar{T},J \bar T$, $T \bar{J}$ Partition Sums From String Theory}}

\bigskip
\centerline{\it Akikazu Hashimoto${}^{1}$ and David Kutasov${}^{2}$}
\bigskip
\smallskip
\centerline{${}^{1}$ Department of Physics,  University of Wisconsin, Madison}
\centerline{1150 University Avenue, Madison, WI 53706, USA}
\smallskip
\centerline{${}^2$ EFI and Department of Physics, University of
Chicago} \centerline{5640 S. Ellis Av., Chicago, IL 60637, USA }

\smallskip

\vglue .3cm

\bigskip

\bigskip
\noindent
We calculate the torus partition sum of a general $CFT_2$ with left and right moving conserved currents $J$ and $\bar J$,  perturbed by a combination of the irrelevant operators $T\bar T$, $J\bar T$ and $T\bar J$. We use string theory techniques to write it as an integral transform of the partition sum of the unperturbed CFT with chemical potentials for the left and right moving conserved charges. The resulting expression transforms in the right way under the modular group, and reproduces the known spectrum of these models.  We also derive a formula for the partition function of deformed $CFT_2$ with non-vanishing chemical potentials.

\bigskip

\Date{7/19}

\newsec{Introduction}

An interesting recent development in Quantum Field Theory (QFT) is the discovery of a class of theories that can be thought of as irrelevant deformations of two dimensional Conformal Field Theories (CFT's), or more generally of Renormalization Group (RG) flows connecting such CFT's. One reason for the interest in these theories is that unlike general irrelevant deformations, they seem to be well defined (in some region in their parameter space). The usual ambiguity in flowing up the RG is eliminated by using symmetries, a mechanism that might be of more general interest. Another reason is that these theories typically have a Hagedorn high energy density of states, \ie\ they do not approach a fixed point of the RG in the UV. The Hagedorn entropy and other considerations suggest that these theories are non-local. 

So far, one of the main results on these theories has been their spectrum on a circle. The original work of \refs{\SmirnovLQW,\CavagliaODA}  was on theories obtained by deforming a CFT (or, more generally, a QFT) by a bilinear in stress-tensors (the so-called $T\bar T$ deformations). That work was generalized to deformations that are products of a conserved $U(1)$ current and a stress-tensor ($J\bar T$ deformations) \refs{\GuicaLIA\ChakrabortyVJA -\ApoloQPQ}, and more recently to a general linear combination of $T\bar T$, $J\bar T$, and $T\bar J$ \refs{\LeFlochRUT,\ChakrabortyMDF}.

Another observable that has been discussed in these theories is the partition sum on the torus. We can parametrize the torus by a complex coordinate 
\eqn\gam{\gamma = \gamma_1 + i \gamma_2,}
with the identifications 
\eqn\gammaid{\gamma \sim \gamma + 2\pi,\;\;\; \gamma\sim \gamma + 2\pi\zeta,  }
where 
\eqn\defzeta{\zeta=\zeta_1+i\zeta_2} 
is the modulus of the torus. The overall scale of the torus is $R$, \ie\ the metric on the torus is
\eqn\mettor{ds^2=R^2d\gamma d\bar\gamma.}
As is standard in QFT, the partition sum of the theory on the torus \gam\ -- \mettor, $\ZZ(\zeta,\bar\zeta, R)$, can be written as
\eqn\ttrr{\ZZ(\zeta,\bar\zeta,R)={\rm Tr} e^{-2\pi\zeta_2R\EE+2\pi i\zeta_1RP},}
where the trace runs over all the eigenstates of the Hamiltonian, $P$ is the momentum of the states, which satisfies the property $n = PR\in Z$, and $\EE$ is the energy of the states.

Of course, once the spectrum is known, the partition sum \ttrr\ can in principle be computed exactly, but it is an interesting observable for a number of reasons. First, one expects on general grounds that it should be modular invariant, if one takes the coupling of the theory to transform appropriately \AharonyBAD. Verifying this transformation property provides a nice check on the spectrum. Second, it is interesting to see the implications of the non-locality of the theory on the behavior of the partition sum, especially in limits in which the torus \gam\ -- \mettor\ becomes small. 

The partition sum \ttrr\ was calculated for the case of $T\bar T$ deformed CFT in \refs{\CardySDV\DubovskyBMO-\DattaTHY,\AharonyBAD}, and for $J\bar T$ deformed CFT in \AharonyICS, and these calculations provided some insights into these theories. It is interesting to perform this calculation for the general case, where all the couplings are present. In this paper we will do this.

As mentioned above, the construction of $T\bar T$, $J\bar T$, and $T\bar J$ deformed CFT relies heavily on symmetries. A related fact is that the spectrum of these theories is universal, in the sense that the energies of the states in the deformed theory depend in a universal way on the energies, momenta and charges of the corresponding states in the undeformed theory, and on the couplings. This feature was systematically exploited in \refs{\AharonyBAD,\AharonyICS}, to study the partition sum of $T\bar T$ and $J\bar T$ deformed CFT's.

Therefore, if we can calculate the partition sum of the general deformed theories for a class of initial CFT's which have the appropriate symmetries, the result of the calculation must be valid in general. In fact, we have access to a large class of CFT's which have precisely the right properties. This class comes from the holographic duality between string theory on $AdS_3$ and certain $CFT_2$'s. It was used in previous work to determine the spectrum of these theories \refs{\GiveonNIE,\GiveonMYJ,\ChakrabortyVJA,\ChakrabortyMDF}. In this paper we will use it to compute their torus partition functions. 

In the next section we briefly review the features of $AdS_3/CFT_2$ and its deformations that will play a role in our discussion, and then turn to the calculation of the torus partition sum.

\newsec{Some aspects of $AdS_3/CFT_2$ and its deformations}

Since this subject has been discussed in detail before (see \eg\ \ChakrabortyMDF), we will be brief here. The class of theories we will be interested in is type II string theory\foot{One can study this construction in the bosonic string as well.} on 
\eqn\aaaa{AdS_3\times\NN,}
where $\NN$ is a compact (worldsheet) CFT. The worldsheet theory on $AdS_3$ has left and right-moving $SL(2,\IR)$ current algebras, with total level $k$. This level determines the size of the anti de-Sitter space, $R_{AdS}=\sqrt k l_s$. The CFT $\NN$ also depends on $k$. For example, its central charge depends on $k$ due to the consistency conditions of string theory.

A prototypical example of the above construction is the near-horizon geometry of $k$ NS5-branes wrapped around $T^4\times S^1$ and $p$ fundamental strings wrapped around the $S^1$, which leads to a background of the form \aaaa, with $\NN=S^3\times T^4$. The level of both $SL(2,\IR)$ and $SU(2)$ current algebras is equal to $k$ (both for left and right-movers), and the string coupling is $g_s^2\sim 1/p$. Note that in this paper we are discussing $AdS_3$ backgrounds \aaaa\ that are supported by NS B-field, under which fundamental strings and NS fivebranes are electrically and magnetically charged, respectively.

The $AdS_3/CFT_2$ duality relates string theory on \aaaa\ to a two dimensional CFT. A lot is known about this duality, but in general, given a choice of $\NN$ it is not known what is the dual CFT. A useful fact for our purposes is that a sector of the boundary CFT is well understood. The sector in question is the long string sector, which we describe next. 

String theory on $AdS_3$ with NS B-field has a spectrum that consists of some discrete states followed by a continuum above a gap. This continuum can be thought of as describing fundamental strings that are moving with arbitrary momentum in the radial direction of $AdS_3$, while wrapping the boundary circle. The long string states are described by the CFT 
\eqn\ddd{\left(\MM_{6k}^{(L)}\right)^p/S_p~,}
where $\MM_{6k}^{(L)}$ is the sigma model
\eqn\aaa{\MM_{6k}^{(L)}=\IR_\phi\times \NN.}
$\IR_\phi$ is a linear dilaton theory, with slope 
\eqn\bbb{Q^{(L)}=(k-1)\sqrt{2\over k}~,}
such that the total central charge of \aaa\ is $c_\MM=6k$.

Note that the statement above concerns the spectrum of the theory. Correlation functions of operators in the theory cannot be computed by using \ddd, \aaa, since they are sensitive to features of the theory in regions other than the ones where \ddd, \aaa\ hold. The situation is similar to that in other non-compact CFT's, such as Liouville theory and the cigar CFT, $SL(2,\IR)/U(1)$, where the spectrum of scattering states can be computed by studying the theory far from a wall, but correlation functions are sensitive to the structure of the wall. 

In order to study $J\bar T$ and $T\bar J$ deformations, we need the compact CFT $\NN$ in the $AdS_3$ background \aaaa\ to contain worldsheet left and right-moving $U(1)$ currents, $K(z)$, $\bar K(\bar z)$ \refs{\GiveonNS,\KutasovXU}. One way to do this, that is sufficient for our purposes, is to take $\NN$ to have the form 
\eqn\compuone{\NN=S^1\times\hat\NN.}
Parametrizing the $S^1$ by a compact coordinate $y$, we have $K(z)=i\partial y$, $\bar K=i\bar\partial y$. Of course, this is a special case of a more general construction. From our perspective, other constructions differ from this one by the spectrum of charges in the undeformed theory. Our main interest is in the dependence of the deformed energies on the undeformed charges (and other quantum numbers), which is universal.

It has been shown in \refs{\GiveonNIE,\ChakrabortyVJA,\ChakrabortyMDF} that deforming the worldsheet theory on $AdS_3\times S^1\times\hat\NN$ by the combination of current bilinears 
\eqn\qqq{\delta\CL_{\rm ws}=\lambda J_{\rm{SL}}^-\bar{J}_{\rm{SL}}^-+\epsilon_+ K\bar{J}_{\rm{SL}}^-+\epsilon_- J_{\rm{SL}}^-\bar K}
is equivalent in the CFT of long strings \ddd, \aaa, to deforming the building block of the symmetric product, $\MM_{6k}^{(L)}$, by adding to its Lagrangian the combination of irrelevant couplings 
\eqn\rrr{\delta\CL_{\rm st}=-\left(t T\bar T+\mu_+ J\bar T+\mu_- T\bar J\right).}
$J_{\rm{SL}}^-$ in \qqq\ is the worldsheet $SL(2,\IR)_L$ current, whose zero mode gives the $SL(2,\IR)_L$ generator $L_{-1}$ in the boundary Virasoro algebra \refs{\GiveonNS,\KutasovXU}. $J(x)$ is the $U(1)$ current in the boundary theory associated with $K(z)$ via the construction of \refs{\GiveonNS,\KutasovXU}. The spacetime couplings $(t,\mu_\pm)$ are proportional to the worldsheet couplings $(\lambda, \epsilon_\pm)$. The precise relation can be obtained, for example, by comparing the results in special cases, such as pure $T\bar T$ and $J\bar T$.\foot{In both the worldsheet theory \qqq\ and the spacetime theory \rrr, there is a freedom of changing the contact terms between the currents ($K(z)$ and $\bar K(\bar z)$ in \qqq, and $J(x)$ and $\bar J(\bar x)$ in \rrr). This  corresponds to the freedom of reparametrization of the space of couplings \KutasovXB, and in particular to the mixing of $\lambda$ with $\epsilon_+\epsilon_-$ in \qqq, and of $t$ with $\mu_+\mu_-$ in \rrr. This freedom needs to be taken into account when comparing the two theories.}

The worldsheet description \qqq\ is useful since current-current (generalized abelian Thirring) deformations are well understood. In \ChakrabortyMDF\ it was used to obtain the spectrum of the deformed theories \rrr\ as a function of the couplings. An important tool in this analysis was the structure of the sigma model obtained by deforming $AdS_3\times S^1\times\hat\NN$ by \qqq. It is given by 
\eqn\edWZW{S(\lambda,\epsilon_+,\epsilon_-)=\frac{k}{2\pi} \int d^2 z\left(\partial\phi\bar{\partial}\phi+h\partial\bar{\gamma}\bar{\partial}\gamma+\frac{2\epsilon_+h}{\sqrt{k}} \partial y\bar{\partial}\gamma+\frac{2\epsilon_-h}{\sqrt{k}}\partial\bar{\gamma}\bar{\partial}y+\frac{f^{-1}h}{k}\partial y\bar{\partial}y \right),}
where
\eqn\harmf{\eqalign{f^{-1}=& \lambda + e^{-2\phi},\cr
 h^{-1}=& \lambda-4\epsilon_+\epsilon_-+e^{-2\phi}.}}
 There is also a (four dimensional) dilaton background,
 \eqn\dilaton{e^{2\Phi} = g_s^2 e^{-2\phi}h~.}
 For $\lambda=\epsilon_\pm=0$, the background \edWZW\ -- \dilaton\ reduces to $AdS_3\times S^1$ (the factor $\hat\NN$ in the background \compuone\ is a spectator under this deformation, and we don't write it in \edWZW). For $\lambda>0$, $\epsilon_\pm=0$, this background, which was referred to in \refs{\GiveonZM,\GiveonNIE} as $\MM_3$, interpolates between a linear dilaton spacetime in the UV (large positive $\phi$), and an $AdS_3$ background obtained by adding to the LST $p$ fundamental strings and going to long distances at large negative $\phi$. It can be interpreted as a vacuum of Little String Theory (LST). For $\lambda=\epsilon_-=0$, $\epsilon_+\not=0$, the background is a null warped $AdS_3$ background, which is dual to $J\bar T$ deformed CFT. It was studied using field theoretic techniques in \refs{\GuicaLIA,\ChakrabortyVJA}, and using the (single trace) holographic description discussed here in \refs{\ChakrabortyVJA,\ApoloQPQ}. 
 
Since long strings live everywhere in the radial direction $\phi$, to compute their spectrum one can send $\phi\to\infty$ in \edWZW. This leads to a free worldsheet theory, whose spectrum can be analyzed using standard methods \ChakrabortyMDF. 

Our purpose here will be to repeat the analysis of \ChakrabortyMDF\ for the torus partition sum. On general grounds, we know that:
\item{(1)} By definition, the partition sum of string theory in the background \edWZW, with the coordinates $(\gamma,\bar\gamma)$ parametrizing a torus \gam\ -- \mettor, gives the trace \ttrr\ over the spectrum found in \ChakrabortyMDF. 
\item{(2)} Since the spectrum computed in \ChakrabortyMDF\ is that of free strings, the calculation we need to do is that on a worldsheet torus (not to be confused with the target space torus of point (1)). 
\item{(3)} The computation described above is guaranteed to give a modular invariant answer.
\item{(4)} Due to universality, we must be able to write the answer in a way that is applicable for deforming an arbitrary CFT by \rrr.

\noindent
In the rest of the paper, we will perform the calculation of the torus partition sum in the background \edWZW. In the next section we start, as a warm-up exercise and to demonstrate the technique, with the case of pure $T\bar T$ deformation (\ie\ the case $\epsilon_\pm=0$ in \qqq, and $\mu_\pm=0$ in \rrr). In that case, we reproduce previously known results from our point of view. In the following section we generalize the calculation to arbitrary $\epsilon_\pm$ \qqq.

\newsec{The torus partition function of $T \bar{T}$ deformed CFT}

As explained in the previous section, for the case $\epsilon_\pm=0$, the long string sector of the boundary CFT, \ddd, \aaa, is holographically dual to string theory on $\MM\times\NN$, where $\NN$ is the compact CFT defined in \aaaa, and $\MM$ is a sigma model on the background 
\eqn\TTS{S = {1\over2\pi}  \int_{\tau} d^2 z \left(\partial \phi \bar \partial \phi+  {1\over\lambda} \partial \bar \gamma  \bar \partial \gamma  \right).}
Note that in eq. \TTS:
\item{(1)} The worldhseet coordinate $z$ parametrizes a torus with modulus $\tau$, \ie\ it is identified as $z\sim z+2\pi$, $z\sim z+2\pi\tau$.
\item{(2)} The integration measure is given by $d^2z=dz_1dz_2$. The area of the torus is $\int_\tau d^2z=(2\pi)^2\tau_2$.
\item{(3)} We have normalized the action slightly differently than in \edWZW, which corresponds to a different normalization of the parameter $\lambda$ in \edWZW\ and \TTS.  

\noindent
To calculate the partition sum of the dual theory on a torus, we identify the boundary coordinates $(\gamma,\bar\gamma)$ as in \gammaid, and compute the partition sum of string theory on $\MM\times\NN$ on a worldsheet torus. The discussion of the previous section implies that the result should give the partition sum of the symmetric product theory \ddd\ --  \bbb, with the building block \aaa\ $T\bar T$ deformed. Via universality, we can read off from it the partition sum of any $T\bar T$ deformed CFT.

Before we start the calculation, a word on normalizations. The normalization of the first term in \TTS\ enters the result in a subtle way. The partition sum of $\phi$, both on the worldsheet torus, and on the target space torus, is proportional to the range of the $\phi$ coordinate which, in the approximation we are using, is infinite. To perform the calculation, one needs to introduce UV and IR cutoffs, and it is easy to see that once this is done, the normalization of $\phi$ enters through an overall multiplicative factor. We will not be careful about this factor here, since this is not necessary for our purposes.

The normalization of the second term in \TTS\  defines the (dimensionless) coupling $\lambda$, the $T\bar T$ coupling in the block of the symmetric product. This definition might differ from other ones by a rescaling, which can be fixed \eg\ by comparing the resulting energies. 

The worldsheet partition sum of string theory on $\MM\times\NN$ can be written as 
\eqn\wspart{Z_{\rm ws}=Z_\gamma Z_{\perp}Z_{\rm gh},
}
where $Z_\gamma$ is the partition sum of the fields $(\gamma,\bar\gamma)$ (and their superpartners), $Z_{\perp}$ is the partition sum on $\IR_\phi\times\NN$, and $Z_{\rm gh}$ is the partition sum of the (super) conformal ghosts. All partition sums in \wspart\ are functions of the modulus of the worldsheet torus, $(\tau,\bar\tau)$, and $Z_{\rm ws}$ needs to be integrated over all inequivalent values of the moduli.

As is standard in string theory, the ghost partition sum, $Z_{\rm gh}$, cancels the contribution of two towers of oscillators of the matter partition sum,  
$Z_\gamma Z_{\perp}$. One can take these towers to be those associated with $(\gamma,\bar\gamma)$. After this is done, the only remaining contribution of $(\gamma,\bar\gamma)$ to the partition sum comes from their zero modes. We can think of the zero mode contribution as a sum of contributions from maps from the worldsheet torus (with modulus $\tau$) to the spacetime one (with modulus $\zeta$). 

Parametrizing the worldsheet torus by the coordinates $(z,\bar z)$, with
\eqn\zsigma{ z =2\pi( \sigma_1 + \tau \sigma_2),  }
and 
\eqn\sig{0 \le \sigma_1, \sigma_2 \le 1,}
these maps are given by \PolchinskiRQ
\eqn\gammasol{
\gamma_1 =2\pi (w_1 \sigma_1 + m_1  \sigma_2)+2\pi \zeta_1 (w_2 \sigma_1 + m_2  \sigma_2), \qquad
\gamma_2 =2\pi \zeta_2 (w_2 \sigma_1 + m_2  \sigma_2).}
Plugging \gammasol\ into \TTS, we find the action
\eqn\Smwmw{S_{\{m_i,w_i\}} =  {\pi \over2\lambda \tau_2} |m_1 - \tau w_1 + \zeta(m_2 - \tau w_2)|^2 . }
Combining all the elements, we find that the partition sum \ttrr\ is given by
\eqn\stringZ{\ZZ(\zeta,\bar \zeta, \lambda) = {\zeta_2 \over 2 \lambda} \sum_{m_i, w_i} \int_{\FF} {d^2 \tau \over \tau_2^2}  e^{-S_{\{m_i,w_i\}}} Z_{\perp}(\tau,\bar{\tau}).}
Here $\FF$ is a fundamental domain of the modular group, and the sum over $(m_i,w_i)$ runs over all integers. The prefactor in front of the sum comes from the Gaussian path integral over $(\gamma,\bar\gamma)$. It can be read off the textbook analysis of \PolchinskiRQ\ (equations (7.3.6), (8.2.11), generalized from the case of $S^1$ to $T^2$ target). One can also determine it by demanding that the integral \stringZ\ can be written as a trace (with integer coefficients) over the spectrum of the deformed theory.

In our discussion so far, we have omitted an important element. The target space coordinates $(\gamma,\bar\gamma)$ live on a torus \gammaid, but string theory on a two-torus has four real moduli, and we have accounted for only three of them, the modulus $\zeta$ \defzeta, and the size $R$ \mettor. The fourth modulus is the NS B-field. Actually, the background \TTS\ already includes a non-zero B-field: 
\eqn\wsgamma{S = {1\over2\pi\lambda}\int  d^2 z\ \partial \bar \gamma \bar \partial \gamma  = {1\over4\pi\lambda}\int  d^2 z\ \left(\left(\partial \bar \gamma \bar \partial \gamma + \partial  \gamma \bar \partial \bar \gamma\right)+\left(\partial \bar \gamma \bar \partial \gamma - \partial  \gamma \bar \partial \bar \gamma\right) \right). }
The B-field is critical, in the sense that the energy of a string winding around the circle (in Lorentzian signature) is zero, with the energy due to winding exactly canceling against the contribution of the B-field. One can add to the action another term, which changes the B-field from this critical value, 
\eqn\extrab{S_b={i  \tilde B \over 2\pi \zeta_2}\int d^2z \left(\partial \bar \gamma \bar \partial \gamma - \partial  \gamma \bar \partial \bar \gamma\right).}
Plugging \gammasol\ into \extrab, we find that the coupling $\tilde B$ gives an additional multiplicative contribution to the r.h.s. of \stringZ, that takes the form $\exp(-2\pi i\tilde BN)$, where
\eqn\Nmwmw{N = w_1 m_2 - w_2 m_1.}
Thus, the partition sum \stringZ, depends on one more coupling, $\ZZ=\ZZ(\zeta,\bar \zeta, \lambda,\tilde B)$, and the dependence on this coupling is periodic,  $\ZZ(\zeta,\bar \zeta, \lambda,\tilde B+1)=\ZZ(\zeta,\bar \zeta, \lambda,\tilde B)$. 

By Fourier transforming $\ZZ$ in $\tilde B$, we can focus on the contribution to the sum in \stringZ\ with a given value of $N$. We focus here on the contribution with $N=1$. Higher values of $N$ are discussed in \HashimotoHQO.

The partition sum $\ZZ(\zeta,\bar \zeta, \lambda,\tilde B)$ is invariant under the modular transformations 
\eqn\modzeta{\zeta\to {a\zeta+b\over c\zeta+d};\;\;\lambda\to {\lambda\over|c\zeta+d|^2}.}
This $SL(2,Z)$ symmetry is part of the $SO(2,2;Z)$ T-duality group of string theory on a two-torus (see \GiveonFU\ for a review). The B-field does not transform under this symmetry. Hence, the Fourier components of the partition sum, which are given by the sum \stringZ\ with fixed $N$ \Nmwmw, are modular invariant as well. 

To recapitulate, we conclude that the partition sum of string theory on $\MM\times\NN$ is given by \stringZ, and we can restrict the sum over the windings $(m_i,w_i)$ to satisfy $N=1$, with $N$ given by \Nmwmw. The dual CFT (before the deformation) is given by \ddd\ -- \bbb, and imposing the constraint $N=1$ in the string theory, corresponds in that CFT to restricting to states in the building block of the symmetric product $\MM_{6k}^{(L)}$ \aaa, \HashimotoHQO.

The partition sum of the CFT $\MM_{6k}^{(L)}$ is given by $Z_\perp$ above. Therefore, using universality, we can  write our final expression for the partition sum of the theory obtained by $T\bar T$ deforming an arbitrary CFT, with partition sum $Z_{\rm cft}$
\eqn\finalstringZ{\ZZ(\zeta,\bar \zeta, \lambda) = {\zeta_2 \over 2 \lambda} \sum_{m_i, w_i|\;N=1} \int_{\FF} {d^2 \tau \over \tau_2^2}  e^{-S_{\{m_i,w_i\}}} Z_{\rm cft}(\tau,\bar{\tau}).}
To simplify the expression \finalstringZ\ further, one can use the fact that $Z_{\rm cft}(\tau,\bar{\tau})$ is modular invariant, and modular transformations act on $(m_1,w_1,m_2,w_2)$ by permuting them, while keeping $N$ \Nmwmw\ fixed. These properties can be used to trade the sum in \finalstringZ\ with an integral over the fundamental domain $\FF$, for an integral over the whole upper half plane, $\HH_+$, with $(m_i,w_i)$ set to a particular value with $N=1$. 

A convenient value to take is 
\eqn\mwmw{m_2=w_1=1, \;\;\; m_1=w_2=0 \ . }
Plugging \mwmw\ into \finalstringZ, and using \Smwmw, we find
\eqn\kernelTT{\ZZ(\zeta, \bar{\zeta},\lambda)={\zeta_2 \over 2 \lambda}
 \int_{{\cal H}_+}{d^2 \tau\over \tau_2^2} e^{-{\pi \over2 \lambda \tau_2} |\tau - \zeta|^2} Z_{\rm cft}(\tau,\bar \tau) \ . }
The expression \kernelTT\ agrees with previous studies of the torus partition function of $T \bar T$ deformed CFT \refs{\CardySDV,\DubovskyBMO}.  For instance, one can obtain it by manipulating eq.~(52) in \DubovskyBMO. However, the origin of this expression, and in particular that of the modulus $\tau$, seems to be different in the two cases. In our case, $\tau$ is the modulus of the worldsheet torus in a holographic description, whereas in  \DubovskyBMO\ holography does not seem to play a role, and $\tau$ arises from the two dimensional (JT) gravitational description of the deformed theory.

Expression \kernelTT\  is also related to the results of \AharonyBAD. Indeed, it gives the partition sum of the deformed theory as a convolution of the undeformed partition sum with the kernel
\eqn\Izzl{
I(\zeta, \bar \zeta, \tau, \bar \tau| \lambda) = 
{\zeta_2 \over 2 \lambda\tau_2^2}
e^{-{\pi \over2 \lambda \tau_2} |\tau - \zeta|^2}.}
This kernel satisfies the differential equation
\eqn\diffeq{ \partial_\lambda I(\zeta, \bar \zeta, \tau, \bar \tau| \lambda) = {\pi \over 2} \left[\zeta_2 \partial_\zeta \partial_{\bar \zeta}+{1 \over 2} \left(i (\partial_\zeta - \partial_{\bar \zeta}) - {1 \over \zeta_2}\right) \lambda \partial_\lambda \right]  I(\zeta, \bar \zeta, \tau, \bar \tau| \lambda)}
for all $\tau$.  As a result, the partition function \kernelTT\ satisfies the same differential equation, in agreement with eq. (3.1) in \AharonyBAD. Since \Izzl\  approaches a $\delta$ function as $\lambda\to 0$, \kernelTT\ satisfies the initial condition (3.2) in \AharonyBAD\ as well. The above discussion can be used to determine the relation between the coupling $\lambda$ defined here, and the one defined in  \AharonyBAD. One finds $\lambda_{\rm here}={\pi\over2}\lambda_{\rm there}$.

We finish this section with a few comments about the expression \kernelTT:
\item{(1)} The integral \kernelTT\ is only convergent for $\lambda>0$, the region in which the model is well defined non-perturbatively \AharonyBAD. For negative $\lambda$, the kernel \Izzl\ is highly non-normalizable, and the integral \kernelTT\ is highly divergent. This is related to the non-perturbative ambiguities found in \AharonyBAD.
\item{(2)} The partition sum of the deformed theory is obtained by smearing that of the undeformed theory over a region whose size grows as the coupling $\lambda$ increases. This is the way that the torus partition sum exhibits the non-locality of the theory, and the fact that the non-locality scale is proportional to $\sqrt\lambda$.
\item{(3)}  It makes the behavior of the partition function $\ZZ$ under modular transformations manifest. Indeed, the r.h.s. of \kernelTT\ is a product of four components, each of which is invariant under the $SL(2,Z)$ transformation
\eqn\ztabcd{\zeta \rightarrow {a \zeta + b \over c \zeta + d}, \;\;\; \tau \rightarrow {a \tau + b \over c \tau + d}~,}
which implies, among other things, 
\eqn\tauminuszeta{(\tau - \zeta) \rightarrow {\tau -\zeta \over (c \zeta + d) (c \tau + d)}~,\;\;\; \tau_2\rightarrow{\tau_2\over|c\tau+d|^2}~,\;\;\;\zeta_2\rightarrow{\zeta_2\over|c\zeta+d|^2}~.}
Thus, the partition function \kernelTT\  satisfies
\eqn\modularz{\ZZ \left({a \zeta+b \over c \zeta+d}, {a \bar{\zeta}+b \over c \bar{\zeta}+d}, {\lambda \over |c \zeta + d|^2}\right)= \ZZ(\zeta, \bar{\zeta},\lambda),}
as expected \AharonyBAD.
\item{(4)} Equation \kernelTT\ can also be used to infer the deformed spectrum in terms of the undeformed one. The partition sum of the undeformed theory can be written as 
\eqn\zcft{Z_{\rm cft}(\tau,\bar{\tau}) = \sum_n e^{-2 \pi \tau_2 RE_n + 2\pi i \tau_1 R P_n}\ . }
Plugging \zcft\ into \kernelTT, and performing the $\tau$ integral using the identity
\eqn\intid{\int_0^\infty {dt \over t^{3\over2}} \ e^{-a/t-bt} = \sqrt{\pi \over a} e^{-2 \sqrt{ab}}\ ,}
we find
\eqn\deformedZ{
\ZZ(\zeta,\bar{\zeta},\lambda) = \sum_n e^{- 2 \pi R\zeta_2 {\cal E}_n(\lambda) + 2\pi i R P_n \zeta_1},}
with
\eqn\deformede{R{\cal E}_n(\lambda) =\sqrt{{1 \over 4 \lambda^2} + {RE_n \over \lambda} + (R P_n)^2}  - {1 \over 2  \lambda} \ , }
in agreement with \AharonyBAD. Note that for vanishing undeformed energy and momentum, one finds that the deformed energy vanishes as well. In the $AdS_3/CFT_2$ context, this state describes a string that winds around the boundary circle without any excitations. The cancellation between the two  terms on the r.h.s. of \deformede\ for it is the cancellation mentioned above between the Nambu-Goto contribution to the energy, and that of the (critical) B-field, see the discussion around \wsgamma.

\newsec{The torus partition function for general couplings}

The discussion of section 3 can be extended to the case of general couplings $(\lambda,\epsilon_\pm)$ in \edWZW. Since this background mixes the deformed field theory base space coordinates $(\gamma,\bar\gamma)$ with the target space coordinate $y$, we need to be more careful about the factor $Z_\perp$ in the worldsheet partition sum \wspart. Using \compuone, we can write it as
\eqn\newzperp{Z_\perp=Z_\phi Z_{\hat\NN} Z_y.}
The first two factors are insensitive to the deformation \edWZW. The last factor is modified, due to the mixing of $y$ with $(\gamma,\bar\gamma)$, but in a way that is familiar from the treatment of the Narain moduli space of string theory compactified on a torus (see \eg\ \GiveonFU). As there, the oscillator contribution to the partition sum of $(\gamma,\bar\gamma,y)$ is not affected (it is given, as usual, by $1/|\eta(\tau)|^6$ for the bosons and the standard expression in terms of $\theta$ functions for the fermions)), and we only need to track the zero mode contribution. This can be done in a very similar way to our discussion of $(\gamma,\bar\gamma)$ in the previous section. 

If $y$ lives on a circle of radius $r$, the analog of \gammasol\ for it is
\eqn\classy{y=2\pi r(w_3\sigma_1+m_3\sigma_2).}
Thus, $\partial y$, $\bar\partial y$ are independent of $z$, like the derivatives of $\gamma$, $\bar\gamma$ in section 3. As there, to evaluate the zero mode contribution to the partition sum, we need to plug \gammasol, \classy\ into the worldsheet action \edWZW, evaluated at large $\phi$, and sum $\exp(-S_{\{m_i,w_i\}})$ over all $(m_i, w_i)$, $i=1,2,3$. 

To see what this procedure gives, one can proceed as follows. The worldsheet Lagrangian for $(\gamma,\bar\gamma, y)$ that follows from  \edWZW\ can be written as
\eqn\genTTS{\LL =  {1 \over 2 \pi} \left[ h (\partial \bar \gamma + 2 \epsilon_+ \partial y) ( \bar \partial \gamma + 2 \epsilon_- \bar \partial y) + \partial y \bar \partial y \right]  }
where 
\eqn\hhhh{h={1\over\lambda - 4 \epsilon_+ \epsilon_-}}
is the value of \harmf\ at $\phi\to\infty$, and again, as in \TTS, we used a different normalization than in \edWZW.

We split the path integral into the contribution of the non-zero modes and that of the zero modes of all the fields. As mentioned above, the only part of the computation that depends on the couplings and moduli is the zero mode contribution of the three fields $(\gamma,\bar\gamma, y)$ in  \genTTS. To decouple them, we introduce the auxiliary complex parameter $\chi$, and rewrite \genTTS\ as 
\eqn\Saux{\LL =    {1 \over 2 \pi} \left[{\bar \chi \chi \over 4 \epsilon_+ \epsilon_- h \tau_2^2}  -{i \bar \chi \over \tau_2} \left(\bar\partial y + {\bar \partial \gamma \over 2 \epsilon_-}\right) -  {i \chi \over  \tau_2} \left(\partial y+ {\partial \bar \gamma \over 2 \epsilon_+} \right) +  \partial y \bar \partial y\right]. }
Note that here $\chi$ is {\it not} a worldsheet field, and the integral over it is a regular (complex) Gaussian integral. This is related to the fact that for the zero modes, all the derivatives of fields in \genTTS, \Saux, are independent of $z$. They do depend on $(m_i,w_i)$ \gammasol, \classy, and as mentioned above, one needs to sum $\exp(-S)$ over all the saddle points (see appendix A).

It is convenient to rewrite \Saux\ as 
\eqn\Sgy{\eqalign{\LL =  & {1 \over 2 \pi } \left[h\partial \bar \gamma \bar \partial \gamma+{1 \over 4 \epsilon_+ \epsilon_- h\tau_2^2} \left(\bar \chi - 2 i \tau_2 \epsilon_- h \partial \bar{\gamma}\right) \left(\chi - 2 i \tau_2 \epsilon_+ h \bar{\partial} \gamma\right) \right]\cr
    +& {1 \over 2 \pi} \left(\partial y \bar \partial y- {i \over  \tau_2} \bar\chi \bar\partial y - {i  \over \tau_2} \chi \partial y\right) . }}
The first line of \Sgy\ is a universal expression, which depends on the couplings and the moduli of the worldsheet and target space tori, but not on the particular CFT that is being deformed. The dependence on the CFT comes from the second line of \Sgy.  It gives rise to the partition sum of the undeformed CFT with chemical potentials for $U(1)_R$ and $U(1)_L$ proportional to $\chi$ and $\bar\chi$, respectively. 

Indeed, comparing the second line of \Sgy\ to (A.14), we see that the path integral over $y$ gives rise in this case to the partition sum 
$Z_{\rm inv}(\tau,\bar \tau, \chi, \bar{\chi})$ defined in (A.7), (A.11), (A.13) (with $\kappa=1$, the appropriate value for this case). Putting this together with the first line, which is evaluated as in section 3, we have
\eqn\stringZchi{\ZZ(\zeta,\bar \zeta, \lambda, \epsilon_+, \epsilon_-) = {\zeta_2 h \over 2 } \int_{\cal F} {d^2 \tau \over \tau_2^2} \int {d^2 \chi  \over \tau_2} \left( {1 \over 2  h \epsilon_+ \epsilon_-}\right)  e^{-S_{\rm inst}} Z_{\rm inv}(\tau,\bar \tau, \chi, \bar{\chi}),}
where 
\eqn\sinstgen{
  S_{\rm inst} =  2 \pi \tau_2 \left(h \partial \bar \gamma \bar \partial \gamma+{1 \over 4 \epsilon_+ \epsilon_- h \tau_2^2} \left(\bar \chi - 2 i\tau_2 \epsilon_- h \partial \bar{\gamma}\right) \left(\chi - 2 i \tau_2 \epsilon_+ h \bar{\partial} \gamma\right)\right).}
Recall that, as in section 3, to evaluate \stringZchi, \sinstgen, we need to plug \gammasol\ into them and perform the sum over $(m_1,m_2;w_1,w_2)$, subject to the constraint $N=1$ \Nmwmw. 

Since the integral \stringZchi\ is modular invariant, we can trade it for an integral over the upper half plane, while restricting the sum over $(m_i,w_i)$ to the single value \mwmw. For this case, one has
\eqn\dbg{\bar \partial  \gamma = - i {(\tau-\zeta) \over 2 \tau_2}, \qquad
 \partial  \bar\gamma =  i {(\bar{\tau}-\bar{\zeta}) \over 2 \tau_2}.}
Plugging \dbg\ into \stringZchi, \sinstgen, we find the final result 
\eqn\genZ{\ZZ(\zeta,\bar \zeta, \lambda, \epsilon_+, \epsilon_-) =  \int_{{\cal H}_+} d^2 \tau  \int_\CC d^2 \chi\  I(\zeta,\bar \zeta, \tau, \bar \tau, \chi, \bar\chi) Z_{\rm inv}(\tau,\bar{\tau}, \chi, \bar \chi),}
where
\eqn\Izzttxx{I(\zeta,\bar \zeta, \tau, \bar \tau, \chi, \bar\chi) = {\zeta_2 \over4   \epsilon_+ \epsilon_- \tau_2^3}
e^{
    - {\pi h |\tau-\zeta|^2 \over 2\tau_2}
    - {\pi\over 2 \epsilon_+ \epsilon_- h \tau_2}
    \left(\bar \chi +  \epsilon_- h (\bar \tau - \bar \zeta) \right)
    \left(\chi -  \epsilon_+ h (\tau - \zeta)\right)} \ . }
Equations \genZ\ and \Izzttxx\ are one of the main results of this paper. One can perform a number of checks on them. 
\item{(1)} It is straightforward to verify that the partition sum is modular invariant, with the couplings transforming as expected
\eqn\Zmodular{\ZZ\left(|c \zeta+d|^2 h,{\epsilon_+ \over c \bar \zeta+d},{\epsilon_- \over c  \zeta+d}, {a \zeta  + b \over c \zeta+d}, {a \bar \zeta + b \over c \bar \zeta+d} \right)=\ZZ(h,\epsilon_+,\epsilon_-,\zeta, \bar \zeta). } 
For $\epsilon_\pm\to 0$, \Zmodular\ reduces to \modularz.
\item{(2)} One can read off from them the spectrum of the theory, as in section 3. One finds
\eqn\gendefE{ R {\cal E} = n -{1 \over 2 A}(B + \sqrt{B^2 - 4 AC}),}
where
\eqn\abc{\eqalign{A &=  \left( (\epsilon_+ + \epsilon_-)^2- \lambda \right), \cr
B & =4 \epsilon_-^2 n+4 \epsilon_- \epsilon_+ n+2 \epsilon_- q_R-2 \epsilon_+ q_L-2 \lambda
   n-1,
 \cr
C & =R E-n-q_R^2+(2 \epsilon_- n+q_R)^2,
    }}
which matches the result in equation (5.31) -- (5.33) of \ChakrabortyMDF\
upon rescaling $\lambda = \lambda^{CGK}/4 $, $\epsilon_+
= \epsilon_+^{CGK}/2$, $\epsilon_-
= \epsilon_-^{CGK}/2$, $q_L = -q_L^{CGK}$, and
$q_R = q_R^{CGK}$. Note that this matching must hold by
construction, due to the fact that in the string theory examples
it must work, and the universality of the spectrum discussed in
section 1.

\item{(3)}The condition $A < 0$ that was shown in \ChakrabortyMDF\ to be necessary and sufficient for the spectrum of the deformed CFT to be sensible, and the dual geometry \edWZW\ to be well behaved, arises in our approach as follows. Starting with \genZ, with $Z_{\rm inv}$ given by (A.7) and (A.13), and performing the $(\chi,\bar \chi,\tau_1)$ integrals, gives rise to an integral over $\tau_2$ of the form \intid\ with an overall  factor $\lambda - (\epsilon_+ + \epsilon_-)^2 \equiv (-A)$
in the denominator of the exponential. Therefore, the width of the
exponential scales as $(-A)$ for small positive  $(-A)$.  This
generalizes the observation in section 3, that $\lambda$ parametrizes the degree of
smearing in \kernelTT, to the case of general $\epsilon_\pm$.

\item{(4)} One can take various limits, such as $\epsilon_\pm\to 0$, which gives back the $T\bar T$ deformed CFT \kernelTT, since the $\chi$ integral localizes to $\chi=0$ in that case. Similarly, the limit $\lambda,\epsilon_-\to 0$ gives rise to $J\bar T$ deformed CFT.

\newsec{Discussion}

In this paper we computed the partition sum of a general $CFT_2$ with holomorphic and anti-holomorphic conserved currents, $J$ and $\bar J$ respectively, perturbed by a general combination of $T\bar T$, $J\bar T$ and $T\bar J$, \rrr. To perform this calculation, we used a technique based on holography, that has been used before to determine the spectra of these theories. Our result, equations \genZ, \Izzttxx, gives the partition sum as an integral transform of the unperturbed CFT. It transforms in the right way  \Zmodular\ under modular transformations, and reproduces the spectrum \gendefE, \abc\ obtained before using the same technique. 

One of the main points of this paper, in addition to the detailed results \genZ, \Izzttxx\ is that to obtain new insights into the deformed CFT's \rrr, it is useful to systematically follow the approach to these theories introduced in \refs{\GiveonNIE, \GiveonMYJ} and followed in later papers. In this approach one uses the large class of examples provided by long strings in $AdS_3$ and their dual $CFT_2$ \ddd, together with the universality discussed in \refs{\AharonyBAD,\AharonyICS}, to obtain properties of any deformed CFT of the form \rrr. It is interesting to follow this logic further. Some examples that one can apply this approach to are the following.

Our results for the partition sum, \kernelTT\ for the $T\bar T$ case and \genZ\ for the general case, were obtained by starting with expressions like \stringZ\ for the $T\bar T$ case, and its analog for the general case, and restricting the sum to $N=1$ \Nmwmw. In the string theory context, this is very natural, as explained in sections 2, 3. It corresponds to the fact that the full CFT we are deforming is a symmetric product \ddd, and the perturbation acts on the building block of this symmetric product. Presumably, summing over all $(m_i,w_i)$ in \stringZ\ should give the partition sum of the full symmetric product of deformed CFT's. In  \HashimotoHQO\  we make this precise, and generalize it to an arbitrary symmetric product CFT.

Another interesting generalization suggested by the string theory construction is to systems that contain multiple $U(1)$ symmetries. In string theory this gives rise to Narain compactifications (see \eg\ \PolchinskiRQ), and the associated $SO(d,d;Z)$ T-duality symmetries \GiveonFU. In our context, the string construction provides a unification between the base space torus \gam\ -- \mettor, and the internal $U(1)$'s, and some non-trivial relations that follow from the underlying stringy symmetries. Via universality, these relations must hold for an arbitrary deformed CFT \rrr.

In the discussion of this paper, we restricted attention to the partition sum of the deformed CFT without chemical potentials for the $U(1)$'s.  The generalization of \genZ\ to the case with chemical potential is discussed in Appendix B, with the final result presented in equation (B.2). 
The comparison to the results of \AharonyICS\ however turns out to be somewhat subtle and is left for future investigation.

\bigskip\bigskip
\noindent{\bf Acknowledgements:}
We thank A. Giveon for discussions and collaboration on related issues, and O. Aharony, S. Chakraborty, S. Datta, S. Dubovsky, Y. Jiang and S. Sethi for discussions. We thank the participants of the Simons center workshop ``$T\bar T$ and other solvable deformations of
QFT's'' for many stimulating discussions. The work of AH is supported
in part by DOE grant DE-SC0017647.  The work of DK is supported in
part by DOE grant DE-SC0009924.  DK thanks the Hebrew University, Tel
Aviv University and the Weizmann Institute for hospitality during part
of this work.

\appendix{A}{Partition sum with chemical potentials in $CFT_2$}

As we saw in the text, the torus partition sum of a $CFT_2$ deformed by a general combination of $T\bar T$, $J\bar T$ and $T\bar J$, \rrr, can be written as an integral transform of the partition sum of the original CFT, with chemical potentials for the charges that couple to $J$ and $\bar J$ turned on. In this appendix we discuss some properties of such partition sums. We start with an example, the CFT of a compact scalar field, which we denote by $y(z)$, and then generalize to other cases. 

We take $y(z)$ to be canonically normalized, 
\eqn\yzlog{y(z) y(0) \sim - \ln|z|^2,}
which corresponds in the textbook analysis of \PolchinskiRQ\ to setting $\alpha'=2$. We also take $y$ to live on a circle of radius $r$, \ie\ identify
\eqn\yperiod{y \sim y + 2 \pi r.}
The resulting CFT has left and right-moving conserved currents 
\eqn\Jy{J(z) = i \partial y(z), \qquad \bar J(\bar z) = i \bar\partial y(\bar z),}
normalized as
\eqn\JJ{J(z) J(0) \sim {1 \over z^2},\;\;\; \bar J(\bar z) \bar J(0) \sim {1 \over \bar z^2}.}
The charges corresponding to \Jy\ are the left and right-moving momenta
\eqn\qLR{p_L = \oint {dz\over2\pi i} \, J(z), \qquad p_R = \oint {d \bar z\over2\pi i}\,    \bar J.}
They take the values (see \eg\ \PolchinskiRQ, eq. (8.2.7)) 
\eqn\qLRy{p_L = {n \over r} + { wr \over 2},\qquad  p_R = {n \over r} - { w r\over 2},}
where $n,w$ are integers, the momentum and winding of a state, respectively.\foot{The charges $p_L$, $p_R$,  \qLR, \qLRy, are denoted by $q_L$, $q_R$ in \abc.}

The partition sum with chemical potentials for $p_L$, $p_R$ is defined by
\eqn\Zy{Z_{\rm cft}(\tau,\bar \tau, \nu, \bar \nu) = {\rm Tr} e^{2 \pi i\tau (L_0-{c\over24})  -2 \pi i \bar \tau(\bar L_0-{c\over24})+ 2 \pi i \nu p_L  - 2 \pi i \bar \nu p_R  } ,}
where $c$ is the central charge of the CFT. For the case of a single scalar field we have $c=1$.

The partition sum $Z_{\rm cft}$ \Zy\ can be computed following standard textbook treatments, such as the one leading to (8.2.9) of \PolchinskiRQ. Using the fact that 
\eqn\lloo{L_0=\half p_L^2+N;\;\;\bar L_0=\half p_R^2+\bar N,}
where $N,\bar N$ are the left and right-moving oscillator levels, one finds
\eqn\Zyans{\eqalign{Z_{\rm cft}(\tau,\bar \tau, \nu, \bar \nu) = &{1\over|\eta(\tau)|^2} \sum_{n,w}
\exp\left[ - 2\pi \tau_2 \left({n^2 \over r^2} + {w^2 r^2 \over 4}\right) + 2 \pi i \tau_1 n w \right. \cr
& \left. + 2 \pi i \nu \left({n \over r} + {wr \over 2}\right)  - 2 \pi i \bar\nu \left({n \over r} - {wr \over 2}\right) \right],
}}
with $\eta(\tau)$ the Dedekind eta function.

To study the transformation properties of \Zyans\ under the modular group, it is convenient to Poisson resum it in the variable $n$. Using (8.2.10) of \PolchinskiRQ, we find
\eqn\ZyPos{\eqalign{Z_{\rm cft}(\tau,\bar \tau, \nu, \bar \nu)= &{r \over \sqrt{2 \tau_2}|\eta(\tau)|^2}\sum_{m,w}
\exp \left[ - {\pi r^2 \over 2 \tau_2}|m - w\tau |^2 \right. \cr
&  \left.
+ {\pi r \nu \over \tau_2} ( m - w\bar \tau)
-{\pi r \bar \nu \over \tau_2}   (m -w \tau) - {\pi \over 2 \tau_2} (\nu - \bar \nu)^2
\right].
}}
It is natural to define the quantity
\eqn\Zinvcft{
Z_{\rm inv}(\tau,\bar \tau, \nu, \bar \nu) =
Z_{\rm cft}(\tau,\bar \tau, \nu, \bar \nu) e^{\pi (\nu - \bar \nu)^2/2 \tau_2},}
which, following the discussion after (8.2.11) of  \PolchinskiRQ, can be shown to be invariant under 
\eqn\transnu{\tau \rightarrow {a \tau + b \over c \tau + d};\;\;\;\nu\to{\nu\over c\tau+d};\;\;\;\bar\nu\to{\bar\nu\over c\bar\tau+d}~.
}
A few comments on \Zinvcft, \transnu:
\item{(1)} These results were derived for the specific case of the compact scalar CFT,  \yzlog, \yperiod, with the chemical potentials $\nu$, $\bar\nu$ coupling to currents normalized as in \JJ, but it is easy to generalize them to a general CFT, and currents normalized in a more general way $J(z) J(0) \sim \kappa / z^2$, $\bar J(\bar z) \bar J(0) \sim \kappa / \bar z^2$. In the general case, the relation \Zinvcft\ is replaced by
\eqn\zinv{ Z_{\rm inv}(\tau,\bar \tau, \nu, \bar{\nu})  = Z_{\rm cft}(\tau, \bar\tau, \nu, \bar{\nu}) e^{\kappa \pi (\nu-\bar{\nu})^2/2 \tau_2},}
and one can show that $Z_{\rm inv}$ \zinv\ is invariant under \transnu.
\item{(2)} For the case $\bar\nu=0$, where both $Z_{\rm cft}$, $Z_{\rm inv}$ are only functions of $(\tau,\bar \tau, \nu)$, one can check that the invariance of $Z_{\rm inv}$ under modular transformations \transnu\ leads to (1.2) of \AharonyICS, with $k_{\rm there}=\kappa_{\rm here}$. 
\item{(3)} One can think of the partition sum $Z_{\rm inv}$ \Zinvcft\ as corresponding to the Lagrangian
\eqn\yaction{\LL={1 \over 2 \pi}  \left(\partial y \bar \partial y  - {i  \over  \tau_2  } \nu   \partial  y - { i   \over   \tau_2 } \bar \nu\bar \partial y \right). }
The terms proportional to $\nu$ and $\bar\nu$ in \yaction\ receive contributions only from the zero modes of the field $y$. Plugging \classy\ in \yaction, integrating $\LL$ over the torus, and summing over all $(m,w)$ (called $(m_3,w_3)$ in \classy) gives precisely the expression \ZyPos, \Zinvcft\ for $Z_{\rm inv}$.

\appendix{B}{Partition sum with chemical potentials}

In the text of this paper we computed the partition sum of a deformed CFT \rrr\ for general values of the couplings. The deformed theory preserves the two $U(1)$ symmetries corresponding to $J$ and $\bar J$, though the deformed currents cease to be (anti) holomorphic for generic values of the couplings. We can couple the two conserved charges to chemical potentials $(\nu,\bar\nu)$,  and compute the partition sum as a function of $(\nu,\bar\nu)$. In this appendix we write the resulting expression.

The basic idea is to replace $Z_{\rm inv}(\tau,\bar{\tau}, \chi, \bar \chi)$ in \genZ\ by 
\eqn\BZcft{Z_{\rm inv}(\tau,\bar{\tau}, \chi, \bar \chi)=Z_{\rm cft}(\tau,\bar \tau, \chi, \bar \chi) e^{\pi\kappa (\chi - \bar \chi)^2/2 \tau_2}\to Z_{\rm cft}(\tau, \bar\tau,\chi+\nu, \bar\chi+\bar\nu) e^{\pi\kappa (\chi - \bar \chi)^2/2 \tau_2}}
If we write the partition sum $Z_{\rm cft}$ as a trace, as in \Zy\ (with the charges denoted by $Q_{L(R)}$ rather than $p_{L(R)}$), and shift $\chi\to\chi+\nu$, the contribution to the trace of each state is multiplied by a factor $\exp\left[2\pi i(\nu Q_L-\bar\nu Q_R)\right]$. Therefore, the quantity 
\eqn\BgenZnu{\ZZ(\zeta,\bar \zeta, \lambda, \epsilon_+, \epsilon_-; \nu, \bar \nu) =  \int_{{\cal H}_+} d^2 \tau  \int_\CC d^2 \chi\  I(\zeta,\bar \zeta, \tau, \bar \tau, \chi, \bar\chi) Z_{\rm cft}(\tau,\bar{\tau}, \chi+\nu, \bar \chi+\bar \nu)e^{\pi (\chi-\bar \chi)^2/2\tau_2}}
can be written as a trace of the form \deformedZ\ (with all couplings turned on), with a chemical potential coupling to the conserved $U(1)$ charges,
\eqn\BZfinal{\ZZ(\zeta,\bar \zeta, \lambda, \epsilon_+, \epsilon_-; \nu, \bar \nu)=\sum_n e^{-2 \pi R \zeta_2 {\cal E}_n + 2 \pi i \zeta_1 R P_n}\times 
e^{2 \pi i \left(\nu Q_{Ln} - \bar \nu  Q_{Rn}\right)} . }
Note that in deriving this result we assumed that the integral over $\tau$ and $\chi$ in \genZ\ can be interchanged with the trace that enters $Z_{\rm cft}$. This assumption was already made in deriving \gendefE\ from \genZ, and appears to be consistent both with the results on the spectrum in \ChakrabortyMDF, and with the convergence properties of the various sums and integrals. 

Note also that the chemical potentials $(\nu,\bar\nu)$ in \BgenZnu\ couple to the undeformed charges of the states (\ie\ their charges in the original CFT). In \AharonyICS\ it was found to be convenient to define the chemical potential w.r.t. a coupling-dependent charge, since this gave a partition sum with good modular properties ((1.4) of \AharonyICS). This was related to the fact that the deformed theory contained a current that remained holomorphic after the deformation. Such a current does not seem to exist for generic value of the couplings, and correspondingly, the partition sum \BgenZnu\ does not seem to have good transformation properties under the modular group.

\listrefs
\end